\documentclass[12pt,epsfig]{article}
\usepackage{graphicx}
\usepackage{epstopdf}
\def\beq{\begin{equation}}
\def\eeq{\end{equation}}
\def\bea{\begin{eqnarray}}
\def\eea{\end{eqnarray}}

\renewcommand{\thefootnote}

\begin{document}
\begin{center}
{\Large \bf \sf QES solutions of a two dimensional  system with quadratic non-linearities
  }
\end{center}
\begin{center} 
\vspace{1.3cm}
{\sf{Bhabani Prasad Mandal$^{a}$ \footnote {e-mails address: bhabani.mandal@gmail.com} , Brijesh Kumar Mourya$^{a}$\footnote{e-mail address: brijeshkumarbhu@gmail.com} and Aman Kumar Singh $^a$\footnote{e-mail address:aman.strgtr@gmail.com }}\\
\bigskip
{ $^{a}$ Department of Physics, Banaras Hindu University, Varanasi -221005, India  }\\


\bigskip
\bigskip

\noindent {\bf Abstract}

}
\end{center}
 We consider a one parameter family of a PT symmetric two dimensional system with quadratic non-linearities.
 Such systems are shown to perform periodic oscillations due to existing centers. We describe
 this system by constructing a non-Hermitian Hamiltonian of a particle with position dependent mass. 
 We further construct a canonical transformation which maps this position dependent mass system 
to a QES system. First few QES levels
 are calculated explicitly by using Bender-Dunne (BD) polynomial method. 

\medskip
\vspace{1in}

\newpage
\section{Introduction}
In the past two decades, several ground breaking works have extended the domain of quantum theories by 
incooperating non-Hermitian (NH)  Hamiltonians that are symmetric under combined parity and time reversal (PT)
 symmetric operation  \cite{ben4}-\cite{book2}  . It has been shown that the fully consistent quantum theories with real energy eigenvalues,
 complete set of orthonormal eigenfunctions with unitary time evolution for such complex
systems are possible by choosing appropriate Hilbert space with modified inner product \cite{met1} -\cite{met5}. The importance of 
non-Hermitian operators in the various branches of theoretical and experimental physics has been realized 
extensively and is being acknowledged rigorously in recent years \cite{new} -\cite{new7}. New developments and excitements are sprouting
everyday in all areas of physics not only in quantum domain such as , PT phase transition \cite{new4,new5,ant5,ptp,ptp1,ant6}- complex scattering \cite{ss}-\cite{cpa9} etc but also in the classical systems
such as optics \cite{opt1}-\cite{opt6}. The field of non-Hermitian operators has been developed enormously during the last two decades and some of the typical works can be found in \cite{pt21} -\cite{ant4}
The investigation in these fields further boosted exponentially when
some of the predictions in complex quantum theory have been verified experimentally\cite{opt3}-\cite{opt6}  and many new windows have been opened up.

 The non-linearities in the context of non-Hermitian quantum theories have been studied extensively  \cite{nls}-\cite{nls8}.In this present work we consider a one parametric family of two dimensional PT symmetric system with quadratic non-linearities. Such systems are shown to perform periodic oscillations and hence are shown to be non-conservative.
We show that such a system can be studied by constructing an appropriate Hamiltonian. We construct
a one particle NH Hamiltonian whose canonical equations in the phase space describe this non-linear system.
This NH Hamiltonian describes a particle with position dependent mass. Due to the presence of position dependent
mass term it is difficult to study this system quantum mechanically with analytical solutions. We construct an appropriate canonical transformation
to map this system to a quasi exactly solvable (QES) system originally developed in \cite{qes} for the usual quantum theories. For discussion of QES theories in non-Hermitain systems see for instance \cite{qes0, ant4} and see \cite{qes1} for  a review on QES systems.  QES levels can explicitly be calculated 
using BD polynomial method \cite{bd} and see for instance \cite{ant4, bpm} for BD polynomials method in Non-hermitian QES systems. The QES levels for the non-linear system are calculated and are shown to be real even though the original Hamiltonian was NH.\\

Now we present the plan of paper. In the next section we describe one parametric family of 2-d 
non-linear system and show that it performs periodic motion by analysing its behaviour near different fixed points. The canonical 
transformation is constructed in section 3 and QES solutions
are discussed in section 4. Section 5 is kept for conclusion and discussion.

\section{Non-linear system in 2-dimension}
We consider one parameter family of 2-d non-linear system described by the equations,
\begin{eqnarray}
\dot x &= & y + g x y \nonumber \\
\dot y &= & 1-2 x^2-\frac {g y^2} {2} 
\label{1}
\end{eqnarray}
where g is a parameter.\\
We observe that this system is symmetric under the transformation $(x,y,t)\longrightarrow (x,-y,-t)$, which is one
 form of a combined PT transformation in two dimension \footnote {In two dimension parity transformation can be defined in 3 
alternative way $(x,y)\rightarrow (x,-y),(x,y)\rightarrow (-x,y) \ \ \mbox{and} (x,y)\rightarrow (y,x) $}. 

The Jacobian matrix for this system is written as
\begin{eqnarray}
J=  \left [ \begin{array}{c c}
 gy & 1+gx \\ 
-4x &-gy 
\end{array}
\right ]
\end{eqnarray}
The eigenvalues of the Jacobian matrix corresponding to different fixed points are tabulated below to discuss the dynamical behavior of the system.

\begin{table}[h]
\centering 
\begin{tabular}{|c| |c| |c| |c| |c|}
\hline
Fixed points $ \rightarrow  $ & $(\frac{1}{\sqrt{2}},0) $ & $ (-\frac{1}{\sqrt{2}},0) $& $(-\frac{1}{g}, \sqrt{\frac{2}{g}(1-\frac{2}{g^2}}) $ & $ (-\frac{1}{g}, -\sqrt{\frac{2}{g}(1-\frac{2}{g^2}}) $\\ 
Eigenvalues $\downarrow $ & &&& \\ \hline
$\lambda_1 $ & $ \sqrt{-2\sqrt{2}(1+\frac{g}{\sqrt{2}}}) $ & $\sqrt{2\sqrt{2}(1-\frac{g}{\sqrt{2}}})  $& $ \sqrt{2g(1-\frac{2}{g^2}}) $&
$\sqrt{2g(1-\frac{2}{g^2}}) $\\ 
&&&& \\ \hline
$\lambda_2 $& $- \sqrt{-2\sqrt{2}(1+\frac{g}{\sqrt{2}}}) $& $ -\sqrt{2\sqrt{2}(1-\frac{g}{\sqrt{2}}})$ & $-\sqrt{2g(1-\frac{2}{g^2}}) $ & $ -\sqrt{2g(1-\frac{2}{g^2}}) $\\
&&&& \\ \hline 
\end{tabular}
\caption{Fixed points and Eigenvalues}
\end{table}
From this table we can conclude the following things about this system.
(i) Both the roots are imaginary for the first fixed point $(\frac{1}{\sqrt{2}},0)$ for $g> -\sqrt{2}$ and the fixed point behaves as a center. On the other hand for $g\le-\sqrt{2}$, both roots are real and one of them is positive and hence the fixed point behaves as saddle point.
(ii)Similarly the  fixed point $(\frac{-1}{\sqrt{2}},0)$  behaves as saddle point for $g<\sqrt{2}$ and as center for $g\ge \sqrt{2}$
(iii) The roots for  3rd and 4th fixed points in the above table are equal and hence both the fixed points will have same behavior. For $g>\sqrt{2} $ these fixed points behave like saddle points and for $0<g<\sqrt{2}$ both roots are imaginary indicating  that the fixed points are centers.


The change in stability behaviour of these fixed points can be analysed through bifurcation
diagram. This diagram is obtained by plotting the eigenvalues with the parameters g. The
bifurcation points are identified as those values of $g$ where the eigenvalues change their
sign. It must be noted that fixed point (i) is always of center type for positive values of $g$, this
is why this fixed point would not bifurcate with respect to $g$. However, fixed points (ii), (iii)
and (iv) will have different natures depending upon the values of $g$. We have identified the
regions where the natures of fixed points are different by varying the parameter $g$. These
regions have been shown for fixed point (ii) in Fig. 1, and for fixed points (iii), and (iv) in Fig.
2. The stability behaviours of fixed points (iii), and (iv) are similar.
The behaviour of this system in the vicinity of fixed point (i) is shown in Fig. 3. Based on linear
stability analysis, we find that this fixed point behaves as a center when perturbed slightly.
From Fig. 1, we see that fixed point (ii) is a saddle point when $g\le 1.415$. This behaviour is
confirmed through Fig. 4.
We note that the fixed points (iii) and (iv) would exist only for $g \ge \sqrt{2}$. However, for  the $ g   $
values chosen in the present analysis for QES solutions in Sec 4, the fixed points (iii) and (iv) would not come into existence.


\vspace{0.1in}
\includegraphics[scale=0.60]{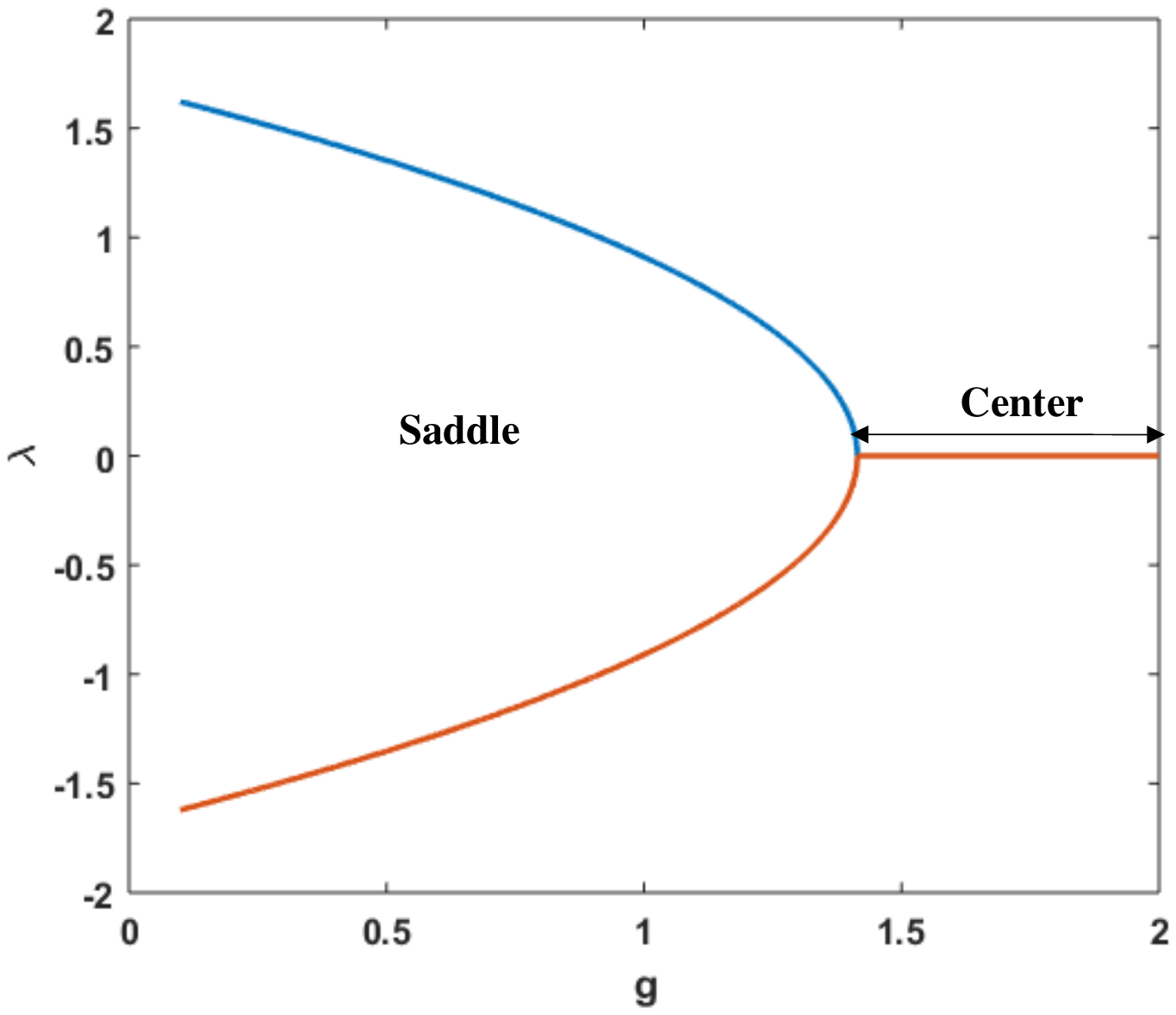}   \\
{\it Fig. 1: The stability behaviour of fixed point (ii). This fixed point behaves as saddle up to
g =1.415. After this value, the equilibrium behaves as a center..}\\ 


\vspace{-0.2in}
 
\includegraphics[scale=0.60]{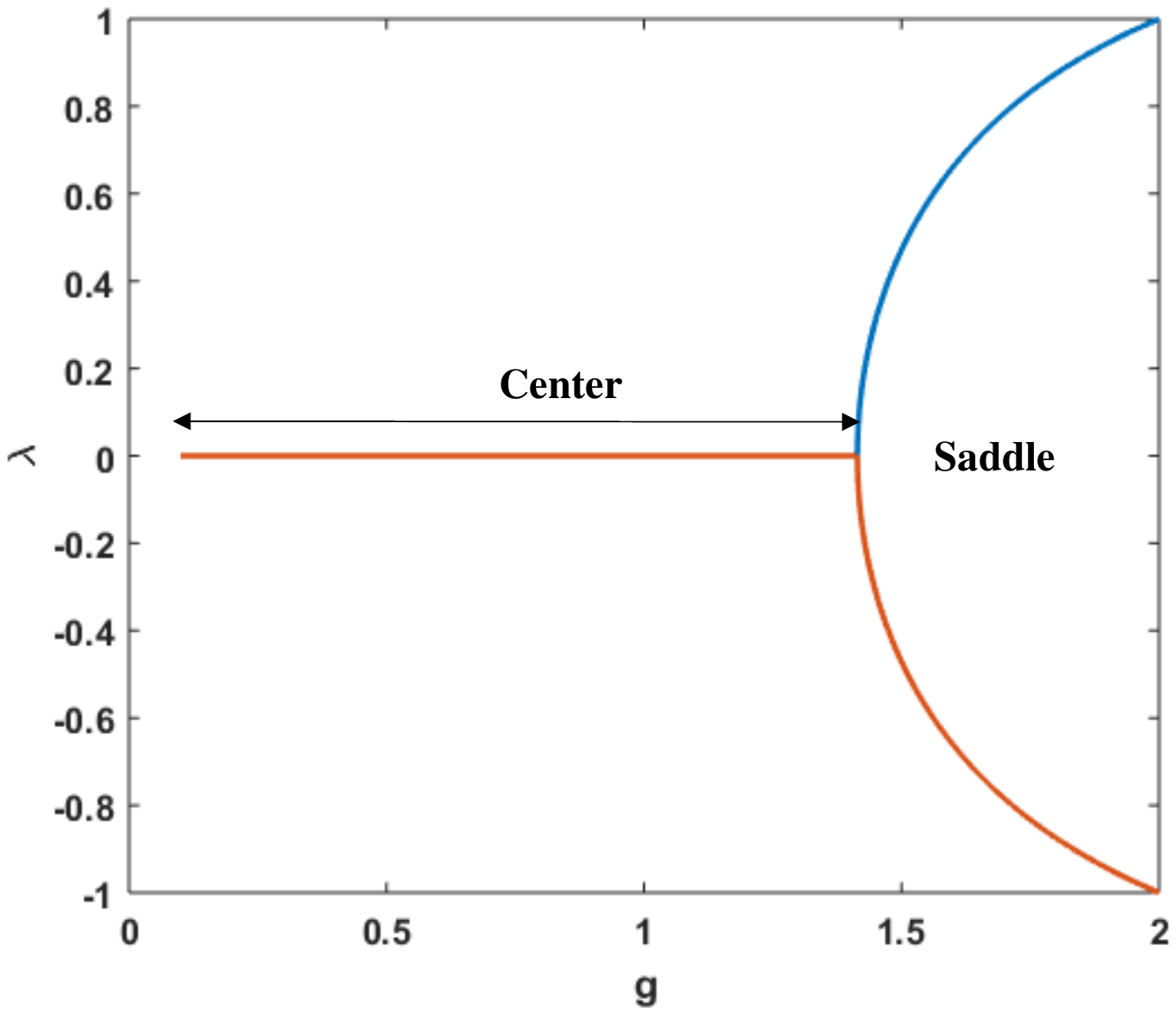}    \\
{\it Fig. 2:The stability behaviour of fixed point (iii) and (iv) are similar. These fixed points
behave as center up to g =1.415. After this value, these fixed points behave as a saddle. }\\ 







\includegraphics[scale=0.70]{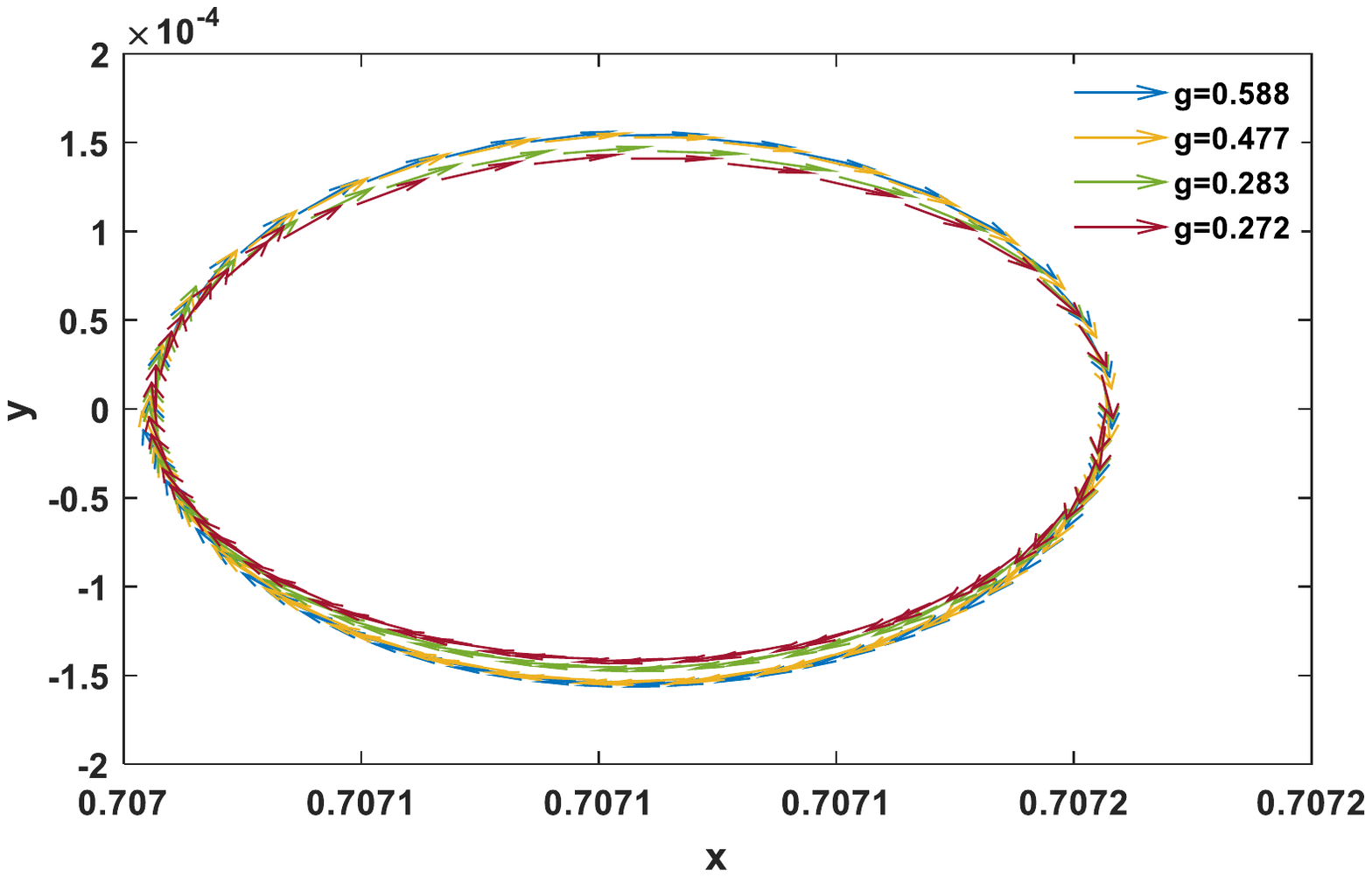}    


{\it Fig. 3: It shows phase space trajectory of non-linear systems. This non-linear system performs periodic motion
for certain range of the parameter g. This is due to the fixed point $(\frac{1}{\sqrt{2}},0 ) $ which behaves like a center. In this parameter range of g the corresponding quantum system admits QES solutions as shown in Sec. 4 } \\ 


\vspace{-0.2in}
 
\includegraphics[scale=0.70]{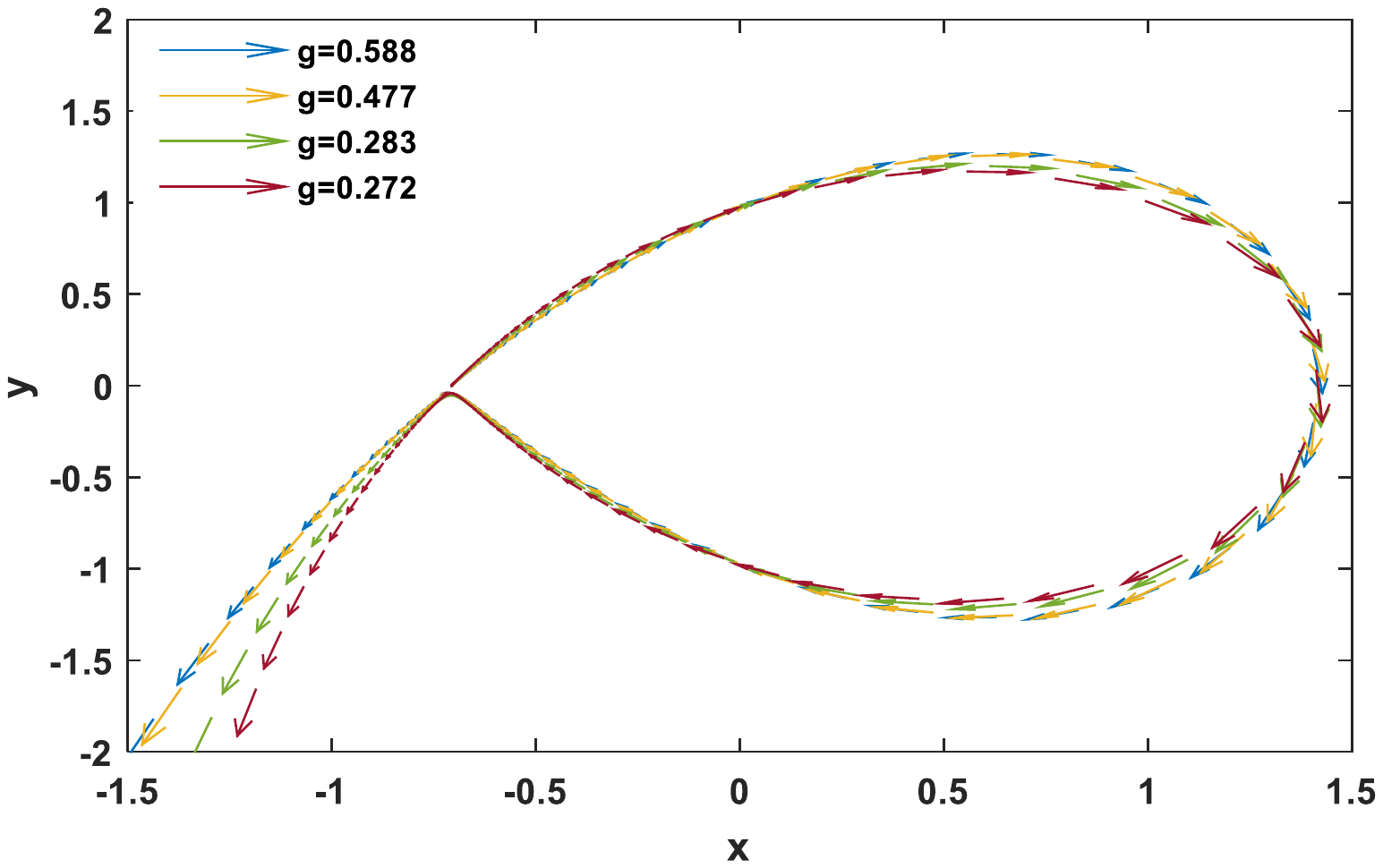}    


{\it Fig. 4: It shows phase space trajectory of non-linear systems.  The trajectories are moving away from the
equilibrium point $(-\frac{1}{\sqrt{2}},0)$This non-linear system performs periodic motion
for certain range of the parameter g.} \\ 

\vspace{0.1cm}

\section{ Canonical transformation}

The non-linear equations in Eq. (\ref{1}) can be obtained through the canonical equation of 1-d Hamiltonian,

\begin{equation} 
H = (1+ g x)\frac{ y^2}{2} + V(x) \equiv (1+ g x)\frac{ p^2}{2} + V(x)
\label{2}
\end{equation}
where $V(x) = \frac{2} {3}  x^3  -  x $, y is treated as canonical conjugate to x, and henceforth is denoted by p.
This Hamiltonian is then interpreted as a single particle systems with position dependent mass.
It is straight forward to see that this system is NH as $ H \neq  H^{\dagger} $. However this 
system is invariant under the transformations $(x,p,t)\longrightarrow (x,-p,-t)$, which is analogous to a
 PT transformation in two dimensional phase space. Due to position 
dependent  mass term it is difficult to find exact spectrum for the system described by H in Eq. (\ref{2}.
In the next section we find QES solution for this system.\\

In this section we construct a canonical transformation which maps the system describe by the Hamiltonian 
in Eq. (\ref{2}) to a QES systems.\\
We construct the transformation 
\begin{eqnarray}
x &= &  \frac{2Q^2-1}{g} \nonumber \\
p &= &  \frac{ g }{4} Q^{-1}P
\label{4}
\end{eqnarray}
One can easily verify that, these are canonical i.e.
\begin{equation}
 \left[Q,P \right] = i \hbar; \quad \mbox {where}  \left[x,p \right] = i \hbar;   
\label{5}
\end{equation}

under this canonical transformation the position dependent Hamiltonian becomes
\begin{equation}
H= \frac{g^2 P^2}{16} + i\hbar\frac{g^2 Q^{-1} P}{16} + V(Q)
\label{6}
\end{equation}
where,
\begin{equation}
V(Q) = \frac{8 a Q^6}{g^3} - \frac{12 a Q^4}{g^3}+ (\frac{6 a }{g^3}-\frac{2b}{g})Q^2+(\frac{-a }{g^3}+\frac{b}{g}) 
\ \ \mbox{with}\ \ a=\frac{2}{3}, b=1.
\label{7}
\end{equation}
Note $ H \not= H^{\dag}$, but this  NH system is PT ($Q \rightarrow -Q, P \rightarrow P, i \rightarrow -i$) symmetric. 
We consider Schroedinger equation $ H \psi = E \psi $ for this system,
\begin{equation}
 \frac{d^2 \psi }{dQ^2}- \frac{1}{Q} \frac{d \psi }{dQ}  +\frac{16}{g^2}[ E - V(Q)]\psi  = 0
\label{8}
\end{equation}
To obtain QES solution we substitute, 
\begin{equation}
 \psi  = e^{-\alpha Q^2- \beta Q^4} \eta (Q)
\label{9}
\end{equation}
in  Eq. (\ref{8}) to obtain
\begin{equation}
 \eta ''(Q)+[-\frac{1}{Q}-4 (\alpha Q + 2 \beta  Q^3)]\eta '(Q)  +[\frac{16 E}{g^2} + \frac{16a}{g^5} -\frac{16b}{g^3}+ (4 \alpha^2-8 \beta -\frac{32}{g^2}(\frac{3a}{g^3} -\frac{b}{g}))Q^2 ]\eta (Q) = 0
\label{10}
\end{equation}
where we have chosen, $\alpha  = - \sqrt{\frac{18a}{g^5}}$ and $\beta   =  \sqrt{\frac{8a}{g^5}}$ ( putting the coeff. of $ Q^4$ and $Q^6 $ as zero).\\

\section{Solutions Using BD Polynomial method}
One of the elegant methods are due to Bender and Dunne\cite{bd}
which uses a new set of orthogonal polynomial in energy variable, E. The main idea here is that the quantum mechanical
wavefunction for a QES systems is the generating functional for the orthogonal polynomial in energy variable $P_n(E)$.
The condition of quasi-exactly solvability is reflected in the vanishing of the norm of all polynomials whose index n exceeds a 
critical polynomials, $P_{j}(E)$. The zeros of the critical polynomial $P_{j}(E)$ are the quasi-exact energy 
eigenvalues of the systems. These polynomials $P_n(E)$, associated with
QES systems satisfied the necessary and sufficient condition for a family of polynomials, with degree  n to form a set of orthogonal 
polynomials with respect to some weight functions $W(E)$ which satisfied 3-term recursion relation of the form 
\begin{equation}
P_n(E) = \left ( A_n E +B_n \right ) P_{n-1}(E) + C_n P_{n-2}(E) \, , \ \ \      
n\ge 1
\label{11}
\end{equation}
where the coefficients $A_n$, $B_n$ and  $C_n$ are  function of E and $A_{n } \not= 0$, $C_1= 0$ and $C_n \not= 0$ for $ n > 1$.\\
For the positive integer value of the parameters j, corresponding to the QES systems, it has been shown that the norm 
of $P_n(E)$ vanishes for $n > j$. All polynomials $P_n(E)$ factor into a product of 
two polynomials, one of which is $P_{j}(E)$,
\begin{equation}
P_{n+j}(E) = P_j(E) Q_n(E)
\label{fac}
\end{equation}
The zeros of the critical polynomials which are the common factor for all higher polynomials $n > J$, are the QES energy 
eigenvalues of quantum mechanical systems. The corresponding eigenfunction for the QES systems are obtained in a straight 
forward manner from these polynomials.

On substituting,
\begin{equation}
 \eta(Q)   = \Sigma \frac {Q^{2n}}{2^{2n}  (n)! (n-1)! } P_n(E)
\label{11}
\end{equation}
in Eq. (9) we obtain the recursion relation satisfied by $P_n(E)$ as
\begin{eqnarray}
P_n(E) +\left [\frac{16 E}{g^2} + \frac{16}{g^2}(\frac{a }{g^3}-\frac{b}{g}) + 24 (n-1)\sqrt{\frac {2a}{g^5}}\right ]P_{n-1}(E)\hspace{1in}\nonumber \\
+\frac {32}{g^5}(n-1)(n-2)[4 b g^2 -3 a -2 (2n-3)\sqrt{2 a g^5}] P_{n-2}(E) = 0
\label{10}
\end{eqnarray}
with initial conditions $P_{-1} =0 $ and $P_0=1$.\\
Let us introduce an arbitrary number J such that 
\begin{equation}
4bg^2-3a = 2 (2J-1)\sqrt{2 a g^5}. 
\end{equation}
 We shall see that when J is a positive
integer, then this represents a QES system. In terms of $J$,
Eq. (\ref{10}) then can be rewritten as 
\begin{eqnarray}
P_n (E)+ \left[ \frac{16 E}{g^2} + \frac{16}{g^2}(\frac{a }{g^3}-\frac{b}{g}) + 24 (n-1)\sqrt{\frac{2a}{g^5}}\right] 
P_{n-1} (E) \hspace{1in}\nonumber \\
-128(n-1)(n-2)(n-J-1) \sqrt{{2a}{g^5}}  P_{n-2} (E) =0 
\label{rec1}
\end{eqnarray}
 from this equation it is clear that when  $J$ is positive integer, this 
recursion relation will reduce to a two term recursion relation. Thus, when $J$ is a positive integer, we have a QES system. \\
These recursion relations generate a set of orthogonal polynomials of which the first few terms are
\begin{eqnarray}
P_1 &=&- \frac{16 E}{g^2} - \frac{16 a}{g^5}+\frac{16 b}{g^3}\nonumber \\  
P_2 &=&  \frac{E^2}{g^4} + [ \frac{2a}{g^5}-\frac{2b}{g^3}+\frac{3}{2}\sqrt{\frac{2a }{g^5}}]\frac{E}{g^2}+(\frac{a} {g^5}-\frac{b} {g^3}+\frac{3}{2}\sqrt{\frac{2a }{g^5}})(\frac{a} {g^5}-\frac{b} {g^3})
 \nonumber \\  &&\hspace{2in} 
\nonumber \\  
\label{pp}
\end{eqnarray}
It is easily seen that when $J$ is a positive integer, exact energy 
eigenvalues for the  
first $J$ levels are known. Further, when $J$ is a positive integer, these
polynomials exhibit 
the factorization property as given by  
\begin{equation}
P_{n+J}(E) = P_J(E) Q_n(E)
\label{fac}
\end{equation}
where the polynomial set $Q_n(E)$ correspond to the non-exact spectrum 
for this problem with $Q_0 (E) =1$. 
 For example, for $J=1$, $P_{n+1}$ will be factorized into $P_1$ and $Q_n$
and the corresponding QES energy level (which in this case is the ground 
state) is obtained by putting
$P_1 =0$ i. e. 
\begin{equation}
E_1= -\frac{a}{g^3} + \frac {b}{g} 
\label{e1}
\end{equation}
Similarly for $J=2$, $P_{n+2}$ will be factorized into $P_2$ and $Q_n$ and
the corresponding energy levels are from $P_2$,
\begin{eqnarray}
E_1 &=& -\frac{a}{g^3} + \frac {b}{g}  
\nonumber \\ 
E_2 &=& -\frac{a}{g^3} + \frac {b}{g} -\frac{3}{2} \sqrt{{2a}{g}} 
 \label{e2}
\end{eqnarray}
For $J=3$, one needs to solve $P_3(E) =0 $ which is a cubic equation in $E$. Hence  we calculate the  QES energy levels upto
$J=10$ numerically which are presented inTable. 2. QES solutions exists for $ g \le 0.58865 $ 

\begin{table}[h]

\centering
\begin{tabular}{|c||c||c|c|}
\hline
S.N  & J & g & E  \\
     & &  \\
\hline
$\{1\}$ &1 & 0.58865 & $P_1(E)=0$, $E_{1}=-0.0816416$  \\
\hline	
$\{2\}$&2 & 0.477122 &$P_2(E)=0$, $E_{1}=-6.54954$, $E_{2}=-4.04201$   \\
\hline 
$\{3\}$&3 & 0.417704&$P_3(E)=0$, $E_{1}=-22.2131$, $E_{2}=-19.0432$ ,$E_{3}=-16.0817$ \\
\hline
$\{4\}$&4 & 0.378671 &$P_4(E)=0$, $E_{1}=-18.3508$, $E_{2}=-15.1846$, $E_{3}=-12.2638$,\\
& & &\ \ \ $E_{4}=-9.63704$   \\
\hline
$\{5\}$&5& 0.350273 & $P_5(E)=0$, $E_{1}=-24.8198$, $E_{2}=-21.4534$ ,$E_{3}=-18.2845$,\\
& & &\ \ \ $E_{4}=-15.3393$, $E_{5}=-12.6578$ \\
\hline
$\{6\}$&6& 0.328305 &$P_6(E)=0$, $E_{1}=-31.5698$, $E_{2}=-28.0402$ ,$E_{3}=-24.6776$,\\
 & & &\ \ \ $E_{4}=-21.4987$, $E_{5}=-18.5262$, $E_{6}=-15.7939$\\
\hline
$\{7\}$&7& 0.310597 & $P_7(E)=0$, $E_{1}=-38.5594$, $E_{2}=-34.8914$ ,$E_{3}=-31.3689$,\\
  & & &\ \ \ $E_{4}=-28.0037$, $E_{5}=-24.8105$, $E_{6}=-21.8095$, \\
 & & &\ \ \                     $E_{7}=-19.0297$\\
\hline
$\{8\}$&8& 0.295893 & $P_8(E)=0$, $E_{1}=-45.7593$, $E_{2}=-41.9705$ ,$E_{3}=-38.3112$,\\
& & &\ \ \ $E_{4}=-34.7902$, $E_{5}=-31.4180$, $E_{6}=-28.2082$, \\
 & & &\ \ \ $E_{7}=-25.1785$, $E_{8}=-22.3542$ \\
\hline
$\{9\}$&9 & 0.283408 &$P_9(E)=0$, $E_{1}=-53.1465$, $E_{2}=-49.2503$ ,$E_{3}=-45.4712$,\\
& & &\ \ \ $E_{4}=-41.8159$, $E_{5}=-38.2924$, $E_{6}=-34.9105$, \\
 & & &\ \ \ $E_{7}=-31.6825$, $E_{8}=28.6244$, $E_{9}=-25.7584$ \\
\hline
$\{10\}$&10 & 0.272619 & $P_{10}(E)=0$, $E_{1}=-60.7037$, $E_{2}=-56.7107$ ,$E_{3}=-38.3112$,\\
& & &\ \ \ $E_{4}=-49.051$, $E_{5}=-45.3961$, $E_{6}=-41.8674$, \\
 & & &\ \ \ $E_{7}=-38.4739$, $E_{8}=-35.2268$, $E_{9}=-32.1407$,\\ 
 & & & \ \ \   $E_{10}=-29.2352$ \\
\hline

\end{tabular}
\caption {Occurrence of  Energy eigenvalues for different values of parameter J and g and for fixed value $a =\frac{2}{3}$, $b=1$.} 
\end{table}

\section {Conclusions}
In this work  a  2-d system with quadratic non-linearities  have been relized  as a PT symmetric  systems in 2-d phase space. We have
constructed an equivalent Hamiltonian whose canonical equation are essentially describe the non-linear systems. This 
Hamiltonian is then represented as one particle systems in 1-d with position dependent mass. It is extremely difficult to solve
such a system. We further constructed an appropriate canonical transformation which maps this systems to a QES system.
Using Bender-Dunne polynomial methods, we explicitly have calculated the first few QES levels explicitly. $ J=1$ to $10$, QES energy levels are calculated numerically and presented in tabular form. We can see from the table 2 that higher $J$ values imply less $g$ values. Maximum $g$ value for which QES solution exists is $g=0.58865$  ( approximately).  In this restriction of g all QES solutions are real, indicating that the system does not possess any exceptional points and the system always remains in the unbroken PT phase. Further note that for the different ranges of the parameter g we have various classical solutions as discussed in Sec. 2, whereas QES quantum solutions exist only for low values of g. The phase space trajectories are drawn ( Figs 3,4 ) in the ranges of g where the corresponding quantum system admits QES solutions. It is interesting to observe that the classical non-linear system performs the regular periodic motion and the first fixed point $(\frac{1}{\sqrt{2}},0)$ behaves as center in the parameter range of   $.272<g <.588 $ (approx) [ See Fig 3] for which the corresponding quantum system admits QES solutions. The  3rd and 4 th fixed points as listed in table 1 also behave as centers in the same parameter range of g.

\end{document}

 \bibitem{qes }A. V. Turbiner, Communications
in Mathematical Physics, 118(3), 467-474]
 
 \bibitem{qes1} A. Fring, (2015) Journal of Physics A: Mathematical
and Theoretical, 48(14), 145301    
       \bibitem{}
    \bibitem{}
    \bibitem{}
    \bibitem{}
  \bibitem{}
    \bibitem{}       
    \bibitem{}
    \bibitem{}
    \bibitem{}
  \bibitem{}
    \bibitem{}      

C. T. West, T. Kottos, and T. Prosen: Phys. Rev. Lett. 104, 054102(2010).
[10] A. Nanayakkara: Phys. Lett. A 304, 67 (2002).
[11] C. M. Bender, G. V. Dunne, P. N. Meisinger, M. Simsek Phys. Lett. A 281 (2001)311-316.

A. Mostafazadeh, Phys. Rev. Lett. 102, 220402 (2009).
[25] Z. H. Musslimani, K. G. Makris, R. El-Ganainy, and D. N. Christodoulides, Phys. Rev. Lett.
100, 030402 (2008).
[26] R. El-Ganainy, K. G. Makris, D. N. Christodoulides and Z. H. Musslimani, Opt. Lett. 32,
2632 (2007).
[27] A. Guo et al, Phys. Rev. Lett. 103, 093902 (2009).
\bibitem{a} J.C. Sprott, X. Wang, G. Chen, Int. J. Bifurc. Chaos Appl. Sci. Eng. {\bf 23}  1350093 (2013).
\bibitem{b} C. Li, J.C. Sprott, Int. J. Bifurc. Chaos Appl. Sci. Eng. {\bf 23}  1350199 (2013).
\bibitem{c} A. Politi, G.L. Oppo, R. Badii, Phys. Rev. A {\bf 4055} (1986).
\bibitem{d} G.A. Leonov, N.V. Kuznetsov, V.I. Vagaitsev, Phys. Lett. A {\bf 375}  2230 (2011).
\bibitem{e} G.A. Leonov, N.V. Kuznetsov, Int. J. Bifurc. Chaos Appl. Sci. Eng. {\bf 23}  12330002 (2013).
\bibitem{f} C. Letellier, L.A. Aguirre, Phys. Rev. E {\bf 85}  036204 (2012).
\bibitem{g} J.C. Sprott, W.G. Hoover, C.G. Hoover,  arXiv:{\bf 1401.1762} [cond-mat.stat-mech.].
\bibitem{ush} A. Ushveridze, {\it Quasi-Exactly Solvable Models
           in Quantum Mechanics}, Inst. of Physics Publishing, Bristol, (1994). 
\bibitem{ush1} A. G. Ushveridze, Mod. Phy. lett {\bf A6}, (977), 1991.
\bibitem{mrt} A. Minzoni, M Rosenbaum and A. Turbiner, hep-th/9606092.
\bibitem{jat}D. P. Jatkar, C. Nagaraja Kumar and A. Khare, Phys. Lett. {\bf A 142},  200, (1989).

\bibitem{cm} F. Calogero and C. Marchioro , J. Math Phys. {\bf 14} , 182, (1973).

\bibitem{mbs} M.V.N. Murthy, R.K. Bhaduri and D. Sen, Phys. Rev. Lett. {\bf 76}, 4103,(1996); 
           R. K. Bhaduri, A. Khare, J. Law, M. V. N. Murthy and D. Sen
            , J. Phys. A : Math. Gen. {\bf 30}  2557, (1997).
\bibitem{cal} F. Calogero, J. Math. Phys. {\bf 10} (1969), 2191, 2197, ibid {\bf 12} (1971) 419.
\bibitem{kuw1} A. Krajewska, A. Ushveridze and Z. Walczak,
               Mod. Phys. Lett. {\bf A 12}, 1225, (1997).
\bibitem{kha} A. Khare, cond-mat/9712133.
\bibitem{chi} T.S. Chihara, {\it An Introduction to Orthogonal Polynomials }
           , Gordon and Breach, New York, (1975); A. Erdelyi, W. Magnus, F. Oberhettinger
              and F. G. Tricomi, {\it Higher Transcendental Functions}, Vol.II, McGraw-Hill, New York, (1953).
\bibitem{fgr}  F. Finkel, A. Gonzalez-Lopez And M. A. Rodriguez, J. Math. Phys. {\bf 37} , 3954, (1996).
\bibitem{kuw} A. Krajewska, A. Ushveridze and Z. Walczak, Mod. Phys. Lett. {\bf A 12} , 1131,(1997).
\bibitem{new} J. Myrheim, E. Halvorsen and A. Vercin, Phys. Lett. {\bf B 278} , 171,(1992).
\bibitem{jk} D. Jatkar and A. Khare, Int. J. Mod. Phys. {\bf A11}  1357,(1996).
\end{thebibliography}

\end{document}